\documentclass[letterpaper, 10 pt, conference]{ieeeconf}
\IEEEoverridecommandlockouts

\title{\LARGE \bf
Model predictive eco--driving control for heavy--duty trucks using Branch and Bound optimization \thanks{This research was partially funded by the European Union’s Horizon 2020 Research and Innovation Programme, Grant Agreement no. 874972, Project LONGRUN.}}

\author{Bart Wingelaar$^{1}$, Gustavo R. Gon\c{c}alves da Silva$^{1}$ and Mircea Lazar$^{1}$%
\thanks{$^{1}$All authors were affiliated with the department of Electrical Engineering, Eindhoven University of Technology, The Netherlands, at the time this research was conducted: 
{\tt\small g.goncalves.da.silva@tue.nl, m.lazar@tue.nl}}
} 

\usepackage{mathtools}
\usepackage{amsmath,amsfonts,amssymb}
\usepackage{xcolor,soul,graphicx,etoolbox}
\usepackage{float}
\usepackage{subfigure}

\definecolor{ggreen}{rgb}{0.0, 0.9, 0.4}

\usepackage[hyphens]{url}

\usepackage[ruled]{algorithm2e}
\SetKwFor{For}{for (}{) $\lbrace$}{$\rbrace$}
\SetArgSty{textnormal}

\AtBeginEnvironment{bmatrix}{\setlength{\arraycolsep}{2.5pt}}
\newcommand{\commentout}[1]{}

\begin{document}
\maketitle

\thispagestyle{plain}
\pagestyle{plain}

\begin{abstract}
Eco--driving (ED) can be used for fuel savings in existing vehicles, requiring only a few hardware modifications. For this technology to be successful in a dynamic environment, ED requires an online real--time implementable policy. In this work, a dedicated Branch and Bound (BnB) model predictive control (MPC) algorithm is proposed to solve the optimization part of an ED optimal control problem. The developed MPC solution for ED is based on the following ingredients. As a prediction model, the velocity dynamics as a function of distance is modeled by a finite number of driving modes and gear positions. Then we formulate an optimization problem that minimizes a cost function with two terms: one penalizing the fuel consumption and one penalizing the trip duration. We exploit contextual elements and use a warm--started solution to make the BnB solver run in real--time. The results are evaluated in numerical simulations on two routes in Israel and France and the long haul cycle of the Vehicle Energy consumption Calculation Tool (VECTO). In comparison with a human driver and a Pontryagin’s Minimum Principle (PMP) solution, 25.8\% and 12.9\% fuel savings, respectively, are achieved on average.
\end{abstract}

\section{INTRODUCTION}
Heavy--duty trucks (HDTs), also known as Heavy--duty Vehicles (HDVs), are responsible for 5\% of the total emissions and a quarter of the road transport emissions in the EU \cite{euemissionswebsite}. Intending to reduce  these emissions for HDVs, the EU has set goals for a 15\% HDV emission reduction in 2025 and a 30\% reduction in 2030 \cite{eu2}. These emission reductions can be obtained as a consequence of reduced fuel consumption, which in turn can be realized by applying eco--driving (ED).

Eco--driving is a driving strategy and its goal is to minimize fuel consumption by determining an optimal velocity profile, which can be defined as an optimal control problem (OCP) \cite{bart}. Using ED is cost--effective \cite{XU2017784} and does not require substantial vehicle hardware modifications \cite{9145839}. This makes ED a promising fuel--saving method for existing and future HDTs with internal combustion engines (ICEs). 

Since HDTs have discrete gear transmissions, the velocity profile optimization is generally a complex mixed--integer nonlinear programming problem. Several approaches are available to solve the ED optimization problem, including analytical solutions to simplified problems \cite{sciarretta}, numerical application of PMP

\cite{nazar,bart,thomassen,ZHU2019562,Hu2016IntegratedOE}, dynamic programming (DP) \cite{Hellstrom2009} and some bi--level optimization methods \cite{9359536,7801904}. The integrality constraint of the gear position can also be relaxed, reducing the problem to a nonlinear programming problem as done in \cite{ROBUSCHI2021109325,ifacpaper}. 
This ED optimization can also run in a model predictive control (MPC) fashion (e.g., \cite{8475029}). In this context, MPC can be particularly useful in cases where spatial and time--varying elements are included in the optimization problem, such as look--ahead information and vehicle--to--everything (V2X) communication \cite{held2020fuel}. However, online MPC inherently requires real--time optimization, which can be challenging \cite{7393561}.

In recent studies \cite{bart,nazar,thomassen,ifacpaper}, this computational burden was reduced by optimization over driving modes. This novel method converts the difficult--to--solve mixed--integer nonlinear programming problem with continuous decision variables (e.g., engine and brake torques) and discrete decision variables (e.g., gears), into an integer programming problem with only discrete decision variables (driving modes). Thus, this formulation converts the optimization problem into a combinatorial optimization problem. Studies based on this approach, \cite{bart,nazar,thomassen,ifacpaper}, apply a numerical PMP solution to achieve real-time implementation. Although a PMP algorithm is computationally efficient, its use can lead to difficulties with constraints handling and co--state initialization.

Therefore, to achieve a MPC solution for ED that is implementable in real-time, it is more of interest to use solvers that are specifically aimed at solving combinatorial problems, such as Branch and Bound (BnB). However, using a general--purpose BnB reintroduces the high computational complexity issue, compromising the real--time constraint. Hence, in this work we investigate and develop a \textit{dedicated} BnB MPC solver for the driving modes eco--driving problem including slope and V2X preview information, such that real--time implementability constraints are met. In relation with previous works \cite{bart, ifacpaper}, due to the inclusion of slope preview, we extend the velocity profile generating capabilities with two extra driving modes: one to accelerate the HDT and one for prolonged negative road gradients. Secondly, we introduce three elements to make the solver real--time implementable: 1) application of context--based (such as cost function, dynamics, constraints, and a look-ahead horizon) node elimination---this element is not present in general--purpose BnB solvers such as BONMIN \cite{bonmin}; 2) the BnB solver is warm--started with a PMP solution \cite{bart}, providing a locally optimal solution; 3) in contrast to general BnB approaches, the lower bound is not formulated with a relaxation of the integrality constraint, instead a custom heuristic based on the same context information as 1) is used.

Furthermore, we include a realistic model for gearbox transmission efficiency to improve fuel consumption assessment. Finally, we evaluate the fuel savings in numerical simulations on three routes: one in Israel, one in France, and the VECTO long haul cycle, and we show that the BnB solver outperforms a human--driver model and a PMP--MPC-based solution.

The remainder of this paper is organized as follows: In section \ref{sec:dyn} we present the longitudinal HDT dynamics. Then in Section \ref{sec:problem}, the ED OCP is formulated for the MPC algorithm, which can be solved with a BnB algorithm (Section \ref{sec:bnb}). In Section \ref{sec:inplementation}, we present the implementation details and results are presented in Section \ref{sec:results}. Finally, conclusions are presented in Section \ref{sec:conclusion}.

\section{Dynamics modeling of heavy--duty trucks}\label{sec:dyn}
In this section we introduce the main modeling components of the HDT and the driving modes.

\subsection{Longitudinal dynamics}
In this work, the longitudinal dynamics is given by modeling the HDT with a positive velocity and considering the following powertrain components: the rear wheels, final drive, discrete gearbox, transmission efficiencies, ICE, (rotational) inertia and a continuous service brake. Moreover, the modeling fundamentals are coherent with the modeling as used in \cite{sciarretta,ifacpaper,nazar,homberg,saerens}.

Since in our problem we are also interested in including spacial elements such as velocity limits, gradient information, and traffic signals and signs, let us first introduce the transformation from time--domain to space--domain $s\in\mathbb{S}$:
\begin{equation}
    \frac{\text{d}v}{\text{d}s}=\frac{\text{d}v}{\text{d}t}\frac{\text{d}t}{\text{d}s}=\frac{\sum F}{m}\frac{1}{v},
\end{equation}
where $m$ is the HDT mass, state $v$ the velocity, and  $\sum F$ the net force acting on the HDT. The equivalent rotational inertia element is included in $m$ since it is independent of the sign of $\frac{\text{d}v}{\text{d}s}$ \cite{forsberg2019}.

Before defining the driving modes, we define three common elements of the system dynamics: resistance forces, engine speed, and transmission efficiencies.
\subsubsection{Resistance forces}
The resistance forces $F_{res}$ are always present and are defined as follows:
\begin{equation}
      F_{res} = \tfrac{1}{2}\rho_ac_dA_fv^2 + mgc_{rr}\,\text{cos}(\alpha) + mg\,\text{sin}(\alpha).
      \label{eq:fres}
\end{equation}
In \eqref{eq:fres}, $\rho_a$ is the air density, $c_d$ the aerodynamic drag coefficient, $A_f$ the  frontal area, $g$ the gravitational constant, and $c_{rr}$ the roll resistance coefficient and $\alpha$ the slope. 
\subsubsection{Transmission efficiencies}
\begin{figure}
    \centering
    \includegraphics[width=\linewidth]{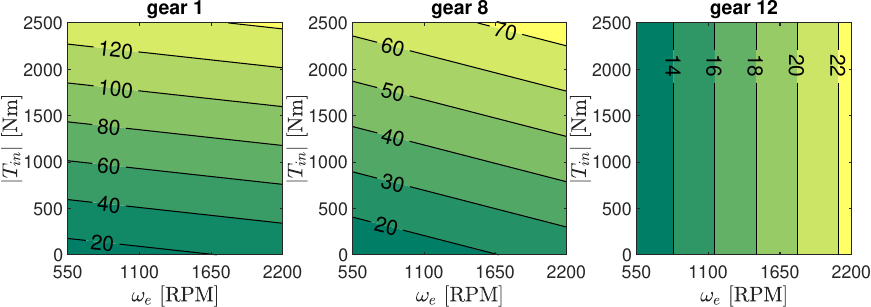}
    \caption{Torque loss maps as function of any transferred torque and engine speed for gears 1,8 and 12.}
    \label{fig:tlossmaps}
\end{figure}

The efficiency of the discrete transmission and final drive ratio, $\eta_t$[-] and $\eta_r$[-] respectively, are described by
\begin{align}
    \eta_t &= 1-\frac{T_{loss}}{|T|}\\
    \eta_r &= c_0,
\end{align}
where $T$ is any torque transferred by the transmission and $c_0$ a constant. $T_{loss}$ is a function of gear ratio, torque and engine speed, and is approximated by a polynomial which is fitted on the data from \cite{vectoFuelEU}:
\begin{equation}
    T_{loss} = c_1(i_t)\omega_e+c_2(i_t)T+c_3(i_t). \label{eq:tloss}
\end{equation}
In \eqref{eq:tloss}, $c_1$--$c_3$ are gear ratio $i_t$--dependent constants. The used torque loss maps are illustrated for gears 1, 8, and 12 in Fig. \ref{fig:tlossmaps}. Note that gear 12 is in a direct drive configuration.
\subsubsection{Engine speed}
Engine speed $\omega_e$ [RPM] is a function of the gear position $y$ [-] and velocity. In neutral gear ($y\!=\!0$), the ICE is disengaged from the drivetrain and idles at a low engine speed.  $\omega_e$ is defined as:
  \begin{equation}
    \omega_e=
    \begin{cases}
          \omega_{idle} & \text{if }y\!=\!0,\\
      \frac{30 i_t(y)i_r}{\pi r_w}\, v & \text{otherwise},\\
    \end{cases}
  \end{equation}
where $r_w$ is the radius of the rear wheels, $i_r$ the final drive ratio and $\omega_{idle}$ the idling speed.

\subsection{Driving modes}
Next, we assume that the velocity profile can be followed by switching between a finite number of subsystems, called driving modes. The resulting switched hybrid system is defined as:
\begin{equation}
       \frac{\text{d}v}{\text{d}s}=f(M(s),v(s)),
\end{equation}
with $f(\cdot)$ the system dynamics of driving mode $M \in\tilde{M}$ with $|\tilde{M}|=6$. The system dynamics is now only dependent on the driving mode as control variable.

\subsubsection{\textbf{Cruising mode $M^{cr}$}}
In cruising mode, the truck drives with a constant velocity and $y>0$. The ICE delivers an engine output torque (as measured by a dynamometer) $T_e$, to overcome $F_{res}>0$ and transmission losses. The dynamics in this mode are described by:
\begin{equation}
   \frac{\text{d}v}{\text{d}s}= \frac{1}{mv}\left(\frac{\eta_r \eta_t i_r i_t(y)}  {r_w}T_e-F_{res}\right)=0.
\end{equation}
\subsubsection{\textbf{Eco--roll mode $M^{ec}$}}
In eco--roll mode, the ICE is disengaged from the drivetrain with the gear position in neutral ($y\!=\!0$) and it rotates at an idling speed $\omega_{idle}$. The HDT dynamics are now fully dependent on $F_{res}$ and the equivalent rotational inertia of the drivetrain $J_{dt}$. The dynamics in this mode are described by:
\begin{equation}
    \frac{\text{d}v}{\text{d}s}=-\frac{r_w^2}{(mr_w^2+J_{dt})v} F_{res}.
\end{equation}

\subsubsection{\textbf{Coasting mode $M^{co}$}}
In coasting mode, the resistance force, equivalent powertrain rotational inertia, and internal engine friction $T_{in.fr}$, determine the HDT dynamics.
The dynamics are described by:
\begin{equation}
        \frac{\text{d}v}{\text{d}s}=-\frac{r_w^2}{(mr_w^2+J_{pt}(y))v}\!\left(\frac{\eta_r \eta_t i_r i_t(y)}  {r_w}T_{in.fr}+F_{res}\right).
\end{equation}
\subsubsection{\textbf{Engine brake mode $M^{eb}$}}
Besides the internal engine friction, an ICE can also deliver a braking force with an engine brake (a type of continuous service brake). In this mode, we assume that maximum engine brake torque $T_{eb,m}$ is applied to decelerate the vehicle.
 The dynamics are described by:
    \begin{equation}
        \frac{\text{d}v}{\text{d}s}\!=\!\frac{-1}{(mr_w^2\!+\!\frac{J_{pt}(y)}{r_w^2})v}\!\left(\!\frac{\eta_r \eta_t i_r i_t(y)}  {r_w}\!(T_{in.fr}\!+\!T_{eb,m})\!+\!F_{res}\!\right)\!,
\end{equation}
where $T_{eb,m}$ is the maximum braking torque of the continuous service brake. The modeling of this brake is based on an approximation of the data in \cite{dafeb}. 
\subsubsection{\textbf{Acceleration mode $M^{ac}$}}
In the acceleration mode, the engine torque is used to accelerate the truck instead of just cruising. The dynamics are described by:
\begin{equation}
    \frac{\text{d}v}{\text{d}s}=\frac{r_w^2}{(mr_w^2\!+\!J_{pt}(y))v} \left(\frac{\eta_r \eta_t i_r i_t(y)}  {r_w}T_{e,ef} - F_{res}\right),\label{eq:dvdsacc}
\end{equation}
where $T_{e,ef}$ is a fixed engine output torque profile based on the optimal brake specific fuel consumption (BSFC) line \cite{forsberg2019}. For a given engine output power $P_e=\frac{30}{\pi}T_e\omega_e$, the corresponding point, ($T_{e,ef}, \omega_e$), on the optimal BSFC line minimizes the following:
    \begin{equation}
        T_{e,ef}(\omega_e) = \min_{T_e} \frac{\dot{m}_f(T_e,\frac{\pi P_e}{30T_e})}{P_e}.
    \end{equation}
This line is illustrated in Fig. \ref{fig:fmap}.

\subsubsection{\textbf{Downhill mode $M^{dh}$}}\label{sec:mdh}
In downhill mode, the HDT maintains a constant speed in negative gradient scenarios. Unlike cruising mode, $F_{res}<0$ due to the negative gradient. Moreover, this mode is only enabled when any gear position in coasting mode cannot provide sufficient braking torque to maintain a constant speed or decelerate the HDT. Therefore, the engine brake provides just enough additional braking torque to maintain a constant speed. The engine brake torque is computed such that:
    \begin{equation}
        \frac{\text{d}v}{\text{d}s}\!=\!\frac{1}{mv}\!\left(\!\frac{\eta_r \eta_t i_r i_t(y)}  {r_w}(T_{in.fr}\!+\!T_{eb})\!+\!F_{res}\!\right)=0.
\end{equation}

\subsection{Fuel consumption}
\begin{figure}
    \centering
    \includegraphics[width=0.9\linewidth]{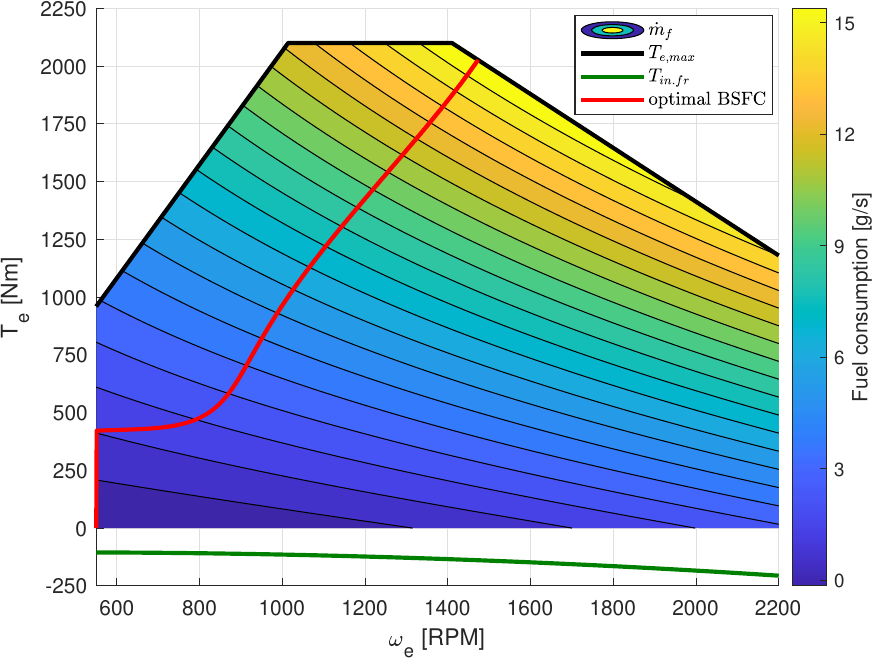}
    \caption{2D polynomial fit of $\dot{m}_f$($T_e$,$\omega_e$) [g/s] (data taken from \cite{ifacpaper}); Optimal BSFC line, maximum $T_e$-constraint and $T_{in.fr}$($\omega_e$)}
    \label{fig:fmap}
\end{figure}

Each driving mode has a specific fuel consumption $\frac{\text{d}m_f}{\text{d}s}$, which is presented in \eqref{eq:mf}:
  \begin{equation}
     \frac{\text{d}m_f}{\text{d}s}=
    \begin{cases}
      \frac{1}{v}\dot{m}_f(T_e,\omega_e) & M\in\{M^{cr},M^{ac}\}\\
               \frac{1}{v}\dot{m}_{f,idle} & M\in\{M^{ec}\}\\
      0 & M\in\{M^{co},M^{eb},M^{dh}\},\\
    \end{cases}
    \label{eq:mf}
  \end{equation}
where $\dot{m}_{f,idle}$ is a small amount of fuel consumed by the idling ICE. The fuel consumption $\dot{m}_f$ is a polynomial fit which maps $T_e$ and $\omega_e$ to the diesel fuel map, given by:
\begin{equation}
    \dot{m}_f = c_4+c_5\omega_e+c_6 T_e+ c_7\omega_e^2+c_8\omega_e T_e+c_9 T_e^2,
\end{equation}
where $c_4$--$c_9$ are constants. In Fig. \ref{fig:fmap}, $\dot{m}_{f}$ is illustrated, as well as the optimal BSFC line, maximum engine output torque constraint $T_{e,max}$, internal engine friction $T_{in.fr}$, and engine speed constraints.  

\section{Formulation of the Eco--driving Optimal Control Problem}\label{sec:problem}
The ED MPC problem formulation can be described by the following OCP:
\begin{equation}
\min_{M\in\tilde{M}} ~ J    
\end{equation}
\begin{equation}
J = \int_{s_0}^{s_f} \left( w_f \frac{\text{d}m_f(M,v)}{\text{d}s} + w_t\frac{1}{v} \right)  
        \text{d}s+\beta(v(s_f)-v_f)^2,\label{eq:ocpct}
\end{equation}
where $J$ [-] is the cost to minimize, $s_0$ the current position, $s_f$ the end position, $w_f\!\in\!(0,1]$ the weight on fuel consumption, $w_t\!\in\![0,1)$ the weight on trip duration, $\beta$ a terminal penalty and $v_f$ an end velocity reference. For appropriate weight scaling, we assume $w_f+w_t=1$.  $w_f$ and $w_f$ can then be adjusted with one parameter, $\phi$ [g/s]:
\begin{equation}
    \phi=\frac{w_t}{w_f}.
\end{equation}
Next, we discretize \eqref{eq:ocpct} in two steps. First, the distance domain $s$ is discretized into $N$ equidistant samples with length $\delta s=\frac{s_f-s_0}{N}$ on which the driving mode--gear combination is constant. Secondly, the state dynamics $f(\cdot)$ are discretized with the forward Euler method. This results in the following discrete OCP at time $k$: 

\begin{subequations}
\label{eq:ocpdt}%
\begin{align}
    &J = \min_{M_{i|k}\in\tilde{M}} \sum_{i=0}^{N-1}L(\Bar{v},M_{i|k})\delta s+\beta(v_{N|k}-v_{f,N})^2\label{eq:c0}\\ 
    &\Bar{v}=\frac{1}{2}(v_{i|k}+v_{i+1|k})\\
    &L(\Bar{v},M_{i|k})=w_f\frac{\text{d}m_f(\!M_{i|k},\Bar{v})}{\text{d}s}+w_t\frac{1}{\Bar{v}}\\
& \text{s.t.}\quad v_{0|k}=v_0\label{eq:c1}\\
  & ~~~~~~~v_{i+1|k}=v_{i|k}+f(M_{i|k},v_{i|k})\delta s \label{eq:c3}\\
    & ~~~~~~~v_{min}(i)\leq v_{i|k}\leq v_{max}(i) \label{eq:c4}\\
  & ~~~~~~~\omega_{e,min}(T_e)\leq\omega_e(y,v_{i|k})\leq\omega_{e,max}(T_e)\label{eq:c5}\\
  & ~~~~~~~ T_e \leq T_{e,max}(\omega_e)\label{eq:c6}\\
    & ~~~~~~~ T_{eb} \leq T_{eb,m}(\omega_e)\label{eq:c7},
\end{align}
\end{subequations}
where $v_0$ is the current velocity of the HDT at time $k$ and \eqref{eq:c3} is the discretized dynamics. \eqref{eq:c4}--\eqref{eq:c7} represent, respectively, constraints on minimum and maximum vehicle speed, engine speed, engine output torque and engine brake torque.

In contrast to conventional ED OCPs, which optimize over discrete and continuous (input) variables ($T$, $T_{eb}$ in \eqref{eq:ocpdt}), the continuous inputs are now fixed by the driving mode dynamics. Consequently, this new formulation is converted into a combinatorial optimization problem, which is typically solved using BnB methods.

\section{Eco--driving MPC Branch and Bound algorithm}\label{sec:bnb}

In this section we introduce the general idea of using MPC with Branch and Bound and the main elements that compose the solution. In Section \ref{sec:inplementation} we then formalize the local elements that make it a dedicated solution.

\subsection{The MPC control part}
\begin{figure}[htb]
    \centering
    \includegraphics[width=1\linewidth]{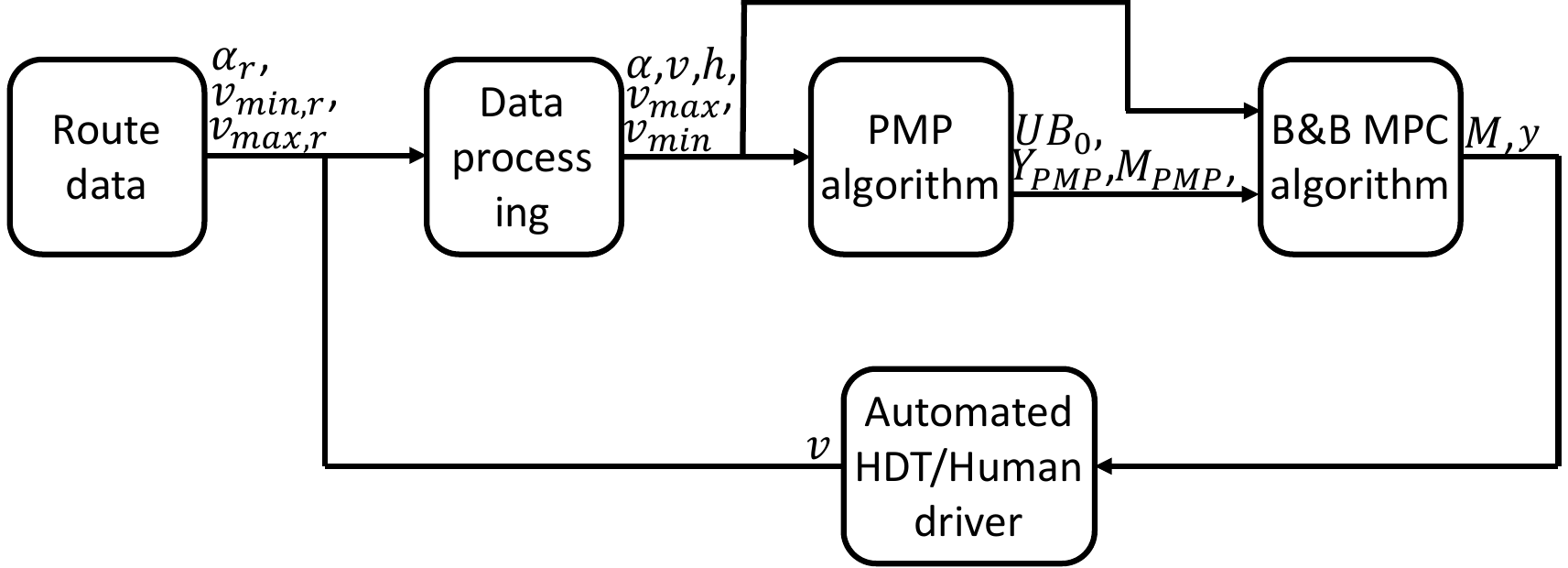}
    \caption{ED MPC algorithm architecture flow chart diagram.}
    \label{fig:edMPCdiagram}
\end{figure}

The MPC part of this algorithm consists of an outer loop where the constraints, cost function and receding horizon are defined. An initial warm--started solution via PMP is also obtained. The BnB algorithm is then used to solve the constrained optimization problem. A schematic overview is provided in Fig. \ref{fig:edMPCdiagram}. 

We assume that live route data is provided to the MPC control loop in the form of $\alpha_r$, $v_{min,r}$ and  $v_{max,r}$. Although temporal events influence this data, for now it is assumed to be constant. Inside the control loop, the incoming data is first pre-processed to extract the relevant data for the MPC OCP ($\alpha$,$v_{min}$,$v_{max}$). Finally, a pseudo--code of the ordered implementation steps of the ED MPC part is provided in Algorithm \ref{alg:edbnbmpcalg}. Notice that the algorithm already includes some context--based steps.
\begin{algorithm}[h]
    \SetAlgoLined
    \textbf{Inputs:} $v_{0|k}$, $v_{min,r}$, $v_{max,r}$, $\alpha$, $\phi$.\\
    1. $[\mathcal{X}_{PMP},{U\!B}_0] = \text{WarmStart}(v_{0|k}, v_{min}, v_{max}, \alpha, \phi)$ // run PMP algorithm and return upper bound and set of nodes $\mathcal{X}_{PMP}$ corresponding to local optimal PMP mode sequence\\
    2. $[v_{min}, v_{max}] = \text{NodeElim1}(v_{min,r}, v_{max,r})$ // update for feasible areas \\
    3. Determine $h$ // determine heuristic term in LB on a regular sized grid for all stage--velocity combinations\\
    4. $\mathcal{X}_{BnB}=\text{EDBnB}(\mathcal{X}_{PMP},{U\!B}_0,v_{0|k}, v_{min}, v_{max}, \alpha, \phi, h)$ // run ED BnB and return set of visited nodes\\
    5. $v^* = \text{OptSeq}(\mathcal{X}_{BnB})$ // determine optimal Node path from visited nodes \\
    6.	$[M^*,y^*] = \text{DetOptPath}(v^*)$ // deduce optimal mode--gear sequence\\
    7. Apply $M[0], y[0]$  // apply first input of optimal sequence\
\caption{ED MPC algorithm}\label{alg:edbnbmpcalg}
\end{algorithm}

\subsection{The BnB algorithm}

Branch and Bound is an algorithm for solving discrete combinatorial optimization  problems, it is generally considered  computationally intensive \cite{Somol2007FastB}, and it can converge to the global optimal solution \cite{Mendona2004OptimizationPI}. 
BnB consists of several fundamental elements: branching,  bounding, a search strategy and elimination. The branching term refers to the exploration of nodes in a tree and the search strategy is the order in which nodes are branched (explored, or visited). This branching operation partitions the problem into smaller sub-problems (nodes) \cite{MORRISON201679}. The search strategy can be categorized in informed search (e.g., best--first search and heuristics) or uninformed search (e.g., depth--first search, breadth--first search) \cite{informedsearch}. 

The bounding term refers to exploring the lower and upper bounds of the nodes. Upper bounds (of which only the smallest is of interest in a minimization problem) indicate a maximum of the domain on which the optimal solution of $J$ can be found. Likewise, the lower bound is a minimum on this domain. In a typical optimization problem, each node has a lower bound, and the best solution thus far is represented by an upper bound. The inclusion of these bounds makes BnB inherently an informed search strategy by itself. As the parent node is branched into child nodes in an iterative way, the lower bounds and upper bounds are evaluated (bounding) for each child node. All child nodes with a lower bound lower than the upper bound have the potential to yield a better solution (a better upper bound). If the lower bound of a node is higher than the upper bound, then branching this node further has no potential to yield a better upper bound. Thus the node is not further explored (elimination or pruning).

In the ED context, the BnB elements have the following meaning. A node corresponds to the HDT velocity at a certain distance stage (or depth level) after having applied a driving mode--gear combination for a certain distance (a multiple of $\delta s$). The stage costs of reaching a node represent the sum of fuel consumption and time duration. Nodes are eliminated when constraints (for instance maximum speed) are exceeded or when the final stage is reached with a lower total cost.

\begin{algorithm}[h]
    \SetAlgoLined
    \textbf{Inputs:} $\mathcal{X}_{PMP},{U\!B}_0,v_{0|k}, v_{min}, v_{max}, \alpha, \phi, h$\\
    1. Set upper bound $U\!B=U\!B_0$\\
    2. Set $\mathcal{X}_{BnB} = \mathcal{X}_{PMP}$\\
    3. Set $\mathcal{B} = v_{0|k}$ and $k=0$\\
    4. Explore nodes until queue is empty or until time
limit is exceed:\\
    \While {$k<N$ AND time$<$time\_limit}{
    4.1 Set $\mathcal{C}=\{ \}$\\
    4.2 \For{$node$ in $\mathcal{B}$}{
        $[\mathcal{C},U\!B] = \text{singleLevelBBSearch}(\mathcal{C},node,h,k,N,U\!B)$ // Algorithm \ref{alg:breadthfirstsearchCustom}
        }  
    4.3 Sort ($\mathcal{C}$) // from lowest LB to highest LB\\
    4.4 $U\!B=\text{GreedyProbing}(U\!B,\mathcal{C}[0],k,N)$ // Algorithm \ref{alg:greedyprobalg}\\
    4.5 Remove nodes in $\mathcal{C}$ with $L\!B\geq U\!B$\\
    4.6 $\mathcal{C} = \text{ContextNodeElim2}(\mathcal{C},v_{min,k},v_{max,k},\varepsilon)$ //  Algorithm \ref{alg:contextNodeElimAlg22} \\
    4.7 $\mathcal{B} = \mathcal{C}$ \\
    4.8 Add $\mathcal{C}$ to $\mathcal{X}_{BnB}$\\
    4.9 $k=k+1$
    }
    \textbf{Output:} $\mathcal{X}_{BnB}$
\caption{ED Branch and Bound algorithm}\label{alg:bnbalg}
\end{algorithm}

A better BnB performance can be achieved by initializing an initial upper bound that is close to the optimal solution. This initial upper bound can be provided by a warm-started feasible solution, such as PMP. Having a feasible solution also means that the BnB algorithm can be terminated at any time. 

Likewise, better lower bounds can also yield better BnB performance and even speed--up computations. These lower bounds are usually composed by a stage cost at stage $j$ and some heuristic term for the cost to go from stage $j$ to the final stage $N$. Its main implication is that it limits the generation of child nodes. In Section \ref{sec:h} we formalize the computation of this bound. 

In Algorithm \ref{alg:bnbalg}, the general idea of the BnB optimization implementation is presented.

\section{Implementation of the BnB MPC algorithm for eco--driving}\label{sec:inplementation}
In this section we describe the elements in MPC and BnB pseudo--codes that make this a dedicated context--based solver. First, we describe node elimination methods based on  exploiting the nature of the problem (this is used in Step 2 of MPC and Step 4.6 of BnB). We then introduce the idea of warm--starting with a fast PMP solver and further we develop an heuristic for the lower--bound computation. Finally we present the two search strategies used with their respective modifications and a sorting algorithm for the lower--bound.

\subsection{Warm start with PMP}\label{sec:PMP}
A BnB algorithm can be warm--started with a feasible initial solution: a feasible sequence of driving modes and the cost thereof as initial upper bound $U\!B_0=J_{PMP}$. In this work, the fast PMP algorithm as developed in \cite{bart} provides that initial solution in case a feasible solution is found ($v_{min}\leq v_{PMP}\leq v_{max}$). If such a solution is found, then the BnB algorithm can be stopped at any time and the final result is feasible. 

However, the PMP algorithm is not guaranteed to yield a feasible solution due to constraint handling difficulties. If no feasible PMP solution is found, $U\!B_0=\infty$. To minimize constraint handling issues, the implementation of the used PMP algorithm is described as follows. First, $v_{max}$ is subdivided into sections where $v_{max}$ is constant. Then the final stage and velocity of each of these sections form a velocity event. Then, the velocity events are solved sequentially with the PMP algorithm. Finally, a terminal penalty is added to the cumulative cost of the individual simulations to formulate the initial upper bound ($J_{PMP}=J_{PMP}+\beta(v_{PMP,N}-v_{f,N})^2$ since the PMP solution solves a two--point boundary value problem without the terminal penalty. 

\subsection{Update maximum and minimum velocities}\label{sec:updatevmax}
With the BnB search tree exploring the distance--velocity domain, it is of interest to exclude infeasible node regions. Avoiding these regions leads to a faster solver. To be more specific, real--world $v_{min}$ and $v_{max}$ are discontinuous and create such regions. The maximum engine power and deceleration power limit the HDT dynamics, which thereby can violate $v_{min}$ and $v_{max}$ constraints if such discontinuities occur. Although the predictive element takes care of this problem and ensures finding a feasible sequence of driving modes, regions of nodes whose child nodes violate these constraints are still explored. 

To overcome this problem, $v_{min}$ and $v_{max}$ were updated such that their derivatives match the maximum acceleration and minimum deceleration of the driving modes. This means that at the beginning of an acceleration event (for example, after a red light), $v_{max}$ is updated forwards as using the acceleration mode, so no node above this line is explored. For a deceleration (so, when reaching a red light), then $v_{max}$ is updated backwards as following engine brake mode. Similarly for $v_{min}$.

\subsection{Lower bound and admissible heuristic}\label{sec:h}
A lower bound is a minimum on the domain on which the optimal solution of $J$ can be found. There are usually no formal closed--form definition of the lower bounds (as they would imply knowledge of the optimal solution), so they are usually composed by a sum of stage costs and some heuristic term. We formalize this definition next. Let us first define the total cost $g$ to reach a node at stage $j$ form stage 0:
\begin{equation}
    g = \sum_{k=0}^{j-1}L_k\label{eq:g}
\end{equation}
where $L_k$ is the stage cost given in (20c).

Next, the lower bound $L\!B$ is formulated as: 
\begin{equation}
  LB=g+h\label{eq:lb},
\end{equation}
where $h$ is a heuristic term that is an estimate on the cost from node $j$ to the final stage $N$. Typically, this heuristic term is formulated by relaxation of the integrality constraint \cite{MORRISON201679}---which turns \eqref{eq:ocpdt} into a nonlinear programming (NLP) problem---, but it can also be simply 0. Furthermore, $h$ must be admissible to retrieve the global optimal solution of $J$. 

In \cite{ifacpaper}, a single case of such an NLP was solved, but this process is not real--time implementable if applied to individual nodes. Moreover, if the resulting NLP solution yields a local optimum, the heuristic is not necessarily admissible and the globally optimal solution is no longer guaranteed to be retrieved. To find a real--time implementable and admissible heuristic, we again use context--specific elements by applying an energy and time--based heuristic similar to \cite{6958015,Ajanovic2017NovelMH}. The added novelty is that for each lower bound, a 2-parameter optimization problem is solved.

Consider node $i$ at stage $j$ and $\Bar{v}_h$ as the average velocity from stage $j$ to stage $N$. The minimum work done by the air drag can then be described by:
\begin{equation}
    W_{air}(\Bar{v}_h)=(N-j)\delta s \frac{1}{2}\rho_a c_d A_f \Bar{v}_h^2.
\end{equation}
This minimum work does not require the velocity profile to be known. The work of the other resistance forces $W_d$ from stage $j$ to stage $N$ is described by:
\begin{equation}
    W_d = \sum_{k=j}^{N-1} mg(c_{rr}\text{cos}(\alpha_k)+\text{sin}(\alpha_k)), 
\end{equation}
which can be determined when slope preview is given. Next, we consider the kinetic energy (excluding rotating parts) of the HDT at the node $i$ with velocity $v_i$ as: 
\begin{equation}
        E_{kin,i}=\frac{1}{2}m v_i^2.
\end{equation}
By determining the energy difference $\Delta E_{kin}$ for going from $j$ to $N$, the energy consumption can be estimated as:
\begin{equation}
    \Delta E_{kin} =  E_{kin,N}-E_{kin,i}+W_d+W_{air}(\Bar{v}_h).
\end{equation}

If $\Delta E_{kin}>0$, then energy is added to the HDT by means of burning fuel. If not, energy is subtracted due to braking forces. With both the energy consumption and time duration defined, we now formulate the total cost--to--go $J_{iN}$ from node $i$ to the final stage: 
\begin{equation}
    J_{iN} = \frac{w_f \text{max}(0,\Delta E_{kin})}{Q\eta_{opt}}+\frac{w_t\delta s (N\!-\!j)}{\Bar{v}_h}\!+\!\beta(v_N\!-\!v_f)^2,
\end{equation}
where $Q$ [J/g] is the specific energy of diesel fuel and $\eta_{opt}$ an estimated maximum combined efficiency of all conversion efficiencies from fuel to work. Notice that the actual value of $J_{iN}$ is not known beforehand, as it depends on the two free variables $\Bar{v}_h$ and $v_N$.
However, for each node, we can solve a small optimization problem with these two variables (all other variables are fixed) in order to provide a minimal admissible value, which is defined as the heuristic $h$:
\begin{equation}
    h(j,v_i) = \min_{\Bar{v}_h,v_N}  J_{iN}(\Bar{v}_h,v_N,j,v_i)\label{eq:minJin}.
\end{equation}

This simple two--variable optimization is solved easily in a (partially) analytic way and is provided to the BnB solver in a 2D lookup table (LUT) with dimensions: node stage and node velocity on a regular sized grid. For each point in the LUT, \eqref{eq:minJin} is calculated and subsequently $h$ can be sampled from the LUT during the BnB optimization.

Both $\Bar{v}_h$, $v_N$ are constraint by $v_{max}$ and $v_{min}$. Moreover, the search domain of $\Bar{v}_h$ can be further reduced since acceleration to $v_{max}$ or deceleration to $v_{min}$ is not instantaneous. Therefore, we update the $v_{max}$ and $v_{min}$ profiles by constraining their derivatives with maximum acceleration and deceleration. To do this efficiently, we assume a maximum ideal acceleration which is defined as applying maximum engine power $P_{e,max}$ to the HDT, defined as:
\begin{equation}
    P_{e,max}=\max \frac{\pi}{30}T_e \omega_e,
\end{equation}
subjected to constraints \eqref{eq:c5} and  \eqref{eq:c6}. Furthermore, the transmission efficiency of the gearbox and equivalent rotational inertia are neglected. The dynamics are described by:
\begin{equation}
    \left(\frac{dv}{ds}\right)_{max} = \frac{\eta_t P_{e,max}-F_{res}v}{mv^2}\label{eq:idealacc}.
\end{equation}
Similarly, the maximum braking power $P_{b,max}$ is defined as:
\begin{equation}
    P_{b,max}=\max \frac{\pi \omega_e}{30}(T_{in.fr}+T_{eb,m}).
\end{equation}
Similarly to \eqref{eq:idealacc}, the minimum ideal deceleration is:
\begin{equation}
    \left(\frac{dv}{ds}\right)_{min} = \frac{-(\eta_t P_{b,max}+F_{res}v)}{mv^2}\label{eq:idealdeacc}.
\end{equation}

\subsection{Search strategy}
Several options exist for the search strategy. For instance, breadth--first search explores all the nodes of a depth level (distance stage) first, before advancing to a deeper level.  Consequently, the upper bound is only updated at the final depth level of the search tree, resulting in many visited nodes which is memory intensive. On the other hand, depth--first search updates the upper bound in the earliest search iterations. However, depth--first search can explore subtrees that do not hold an optimal solution. In this regard, informed search strategies could be a better alternative, although informed search strategies could still end up in an exhaustive search with a large number of visited nodes (e.g., when many nodes have a similar lower bound).

\begin{algorithm}[h]
   \SetAlgoLined
    \textbf{Inputs:} $\mathcal{C},node,h,k,N,U\!B$\\
    1. Branch $node$ in $ChildNodes$ // applying driving mode--gear combinations\\
    2. \For{$childNode$ in $ChildNodes$}{
        $L\!B = \text{DetermineLB}(childNode,h,k)$ //determine $g$ and $L\!B$ using (25) and (26). \\
    2.1 \If {$L\!B<U\!B$ AND $childNode$ is feasible}{
        Add $childNode$ to $\mathcal{C}$\\
        \If{$k==N-1$}{
            $U\!B=L\!B$
            }
        }
    }
\textbf{Output:} $\mathcal{C},U\!B$\\    
\caption{Single level bounded breadth--first search}\label{alg:breadthfirstsearchCustom}
\end{algorithm}
Therefore it is of interest to utilize the strengths of both informed and uninformed search strategies. To this extend, we formulate the search strategy for the ED BnB algorithm as follows. We apply a custom breadth--first search strategy (Algorithm \ref{alg:breadthfirstsearchCustom}) which, after each branched depth level, probes the best child node (node with the lowest lower bound) with a greedy depth policy (Algorithm \ref{alg:greedyprobalg}). This combination has shown to perform well with context--based node elimination (Section \ref{sec:nodelim}). 

The late upper bound update that inherently comes with pure breadth-first search means that few nodes are eliminated based on their lower bounds. This problem is magnified in case of a too low $h$ (see \eqref{eq:lb}). To overcome this, an additional depth probing strategy in combination with a greedy policy was used. A probing strategy keeps a certain branching policy constant and explores the consequences thereof \cite{prob}. Only the best node (node with lowest $L\!B$) is depth probed at each stage after breadth-first branching. This probing is then done with a greedy policy, where only the child node with the lowest $L\!B$ is branched further. This way, the upper bound can be efficiently improved at the earliest stages yielding less node exploration. Finally, it is important to mention that the heuristic term in the lower bound plays an important role in the performance of these search strategies.
\begin{algorithm}[h]
    \SetAlgoLined
    \textbf{Inputs:} $U\!B, \mathcal{D},k,N$\\
    2. \While {$k<N$}{
    2.1 branch $\mathcal{D}$ into $chilNodes$\\
    2.2 $\mathcal{E}$ is $childNode$ with lowest $L\!B$\\
    \uIf{$\mathcal{E}==\O$ or $L\!B$ of $\mathcal{E}>U\!B$}{
           2.3 return // no $U\!B$ update\\
            }
      \Else{
       2.4 $\mathcal{D}=\mathcal{E}$
      }
   2.5 $k=k+1$\\
    }
   3. $U\!B$ = min($U\!B$, $L\!B$ of $\mathcal{E}$)\\
    \textbf{Output:} $U\!B$\\ 
\caption{Greedy probing algorithm}\label{alg:greedyprobalg}
\end{algorithm}

\subsection{ED context--based node elimination}\label{sec:nodelim}

Finally, the BnB problem formulation tends to have an exponential growth in the amount of explored node regardless of the constraints and tight lower bounds. Given the nature of the ED problem, this exponential growth can be avoided. For example, consider two nodes at the same distance stage (or tree depth level) with an almost identical velocity. The cost of reaching these nodes from the root node can significantly deviate. However, the minimum cost of reaching the final distance stage from those two nodes is approximately the same. Therefore it makes no sense to branch both nodes into sub--trees as it has negligible benefit. Instead, only the node with the lowest $g$ (see \eqref{eq:g}) can be branched.

\begin{figure}
    \centering
    \includegraphics[width=0.9\linewidth]{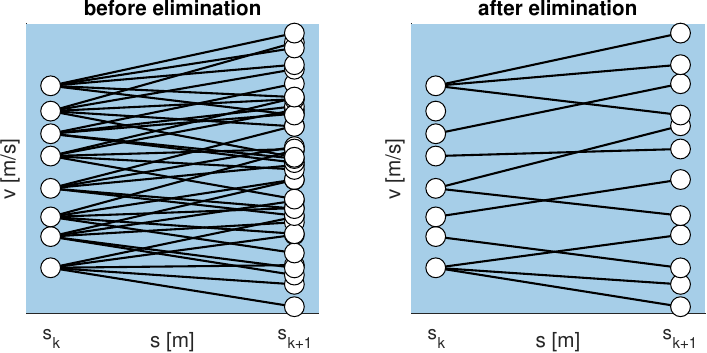}
    \caption{Node elimination process. (left) nodes before context base node elimination; (right) nodes after context based node elimination.}
    \label{fig:nodeElim}
\end{figure}

We applied this procedure as follows. For a given stage, the velocity range $v_{max}-v_{min}$ is discretized into small domains of width $\varepsilon$ ($\varepsilon=0.01$ was used). Only the node with the lowest $g$ in that domain is kept while other nodes with a higher $g$ are eliminated. This process is illustrated in Fig. \ref{fig:nodeElim} and presented in Algorithm \ref{alg:contextNodeElimAlg22}.  
This node elimination is an important factor for making the ED MPC algorithm real--time implementable. It effectively assumes that different nodes in the search tree have the same physical meaning.  Although this method is problem specific, it can be applied to general problems where the nodes have a contextual relation. 
\begin{algorithm}[h]
    \SetAlgoLined
    \textbf{Inputs:} $\mathcal{A},v_{min,k},v_{min,k},\varepsilon$\\
    1. Discretize $v$ into $n$ sections, yielding: $v_{min,k},v_{min,k}\!+\!\varepsilon,v_{min,k},\dots,v_{max,k}\!-\!\varepsilon,v_{max,k}$    \\
    2. Define nodes set $\mathcal{C}$\\
    
     \For{$i = 0;\ i < n;\ i = i + 1$}{
      3.1 $\mathcal{B}=\{nodes\,|\,v_{min,k}+i\varepsilon<nodes\leq v_{min,k}+(i+1)\varepsilon\}\subseteq\mathcal{A}$\\
        3.2 add $node$ with lowest $L\!B$ of $\mathcal{B}$ to $\mathcal{C}$\\
    }
\textbf{Output:} $\mathcal{C}$\\    
\caption{Context based node elimination 2/2}\label{alg:contextNodeElimAlg22}
\end{algorithm}

\section{Results and evaluation}\label{sec:results}

In this section we assess the results obtained with the proposed ED MPC BnB algorithm against a human driver behavior control baseline and an MPC PMP solution for three different routes. The results are evaluated based on: total fuel consumption, trip duration, and the real--time computation constraint. We emphasize a comparison with the PMP solution since a comparison with a human driver has already been presented in \cite{bart}.

Three routes are assessed to avoid making route--specific conclusions. An overview of these routes is presented in Table \ref{tab:routes}. The Israel route is near Neve Zohar, and the France route is from Paris to Rouen. Lastly, the third route is the standardized VECTO long hauling cycle \cite{euvecto}. The $v_{max}$ profile of the routes are shaped to include stop signs, curvatures, and traffic lights events, while $v_{min}=10$~[km/h].

For standstill events we also set $v_{max}=10$~[km/h] for the following reasons: 1) the dynamics discretization (node spacing) is too coarse for the lowest velocities, yielding less feasible nodes; 2) the lowest velocities are incompatible with the $\omega_{e,min}$ constraint; and 3) the dynamics are infeasible at $v=0$. The results in this section are focused on the Israel route due to space limitations.

\setlength{\tabcolsep}{0.2em}
\begin{table}[h]
\centering
\caption{Details of simulated routes}\label{tab:routes}
\begin{tabular}{@{}cccc@{}}
\hline
\textbf{Route}           & \textbf{start coordinate} & \textbf{end coordinate} & \textbf{length[m]} \\ \hline
Israel route    &  30°58'09.6''N & 31°17'25.8"N & 45900 \\ 
                &  35°18'26.2"E  & 35°22'59.8"E & \\ 
France route    &  48°51'25.7"N  & 49°26'35.8"N & 132940 \\
                &  2°20'29.0"E   & 1°05'58.8"E &  \\ 
VECTO cycle & --  & -- & 100180 \\ \hline
\end{tabular}
\end{table}

\subsection{Baseline reference of human driver}
\begin{figure*}[t!]
    \centering
    \includegraphics[width=\linewidth]{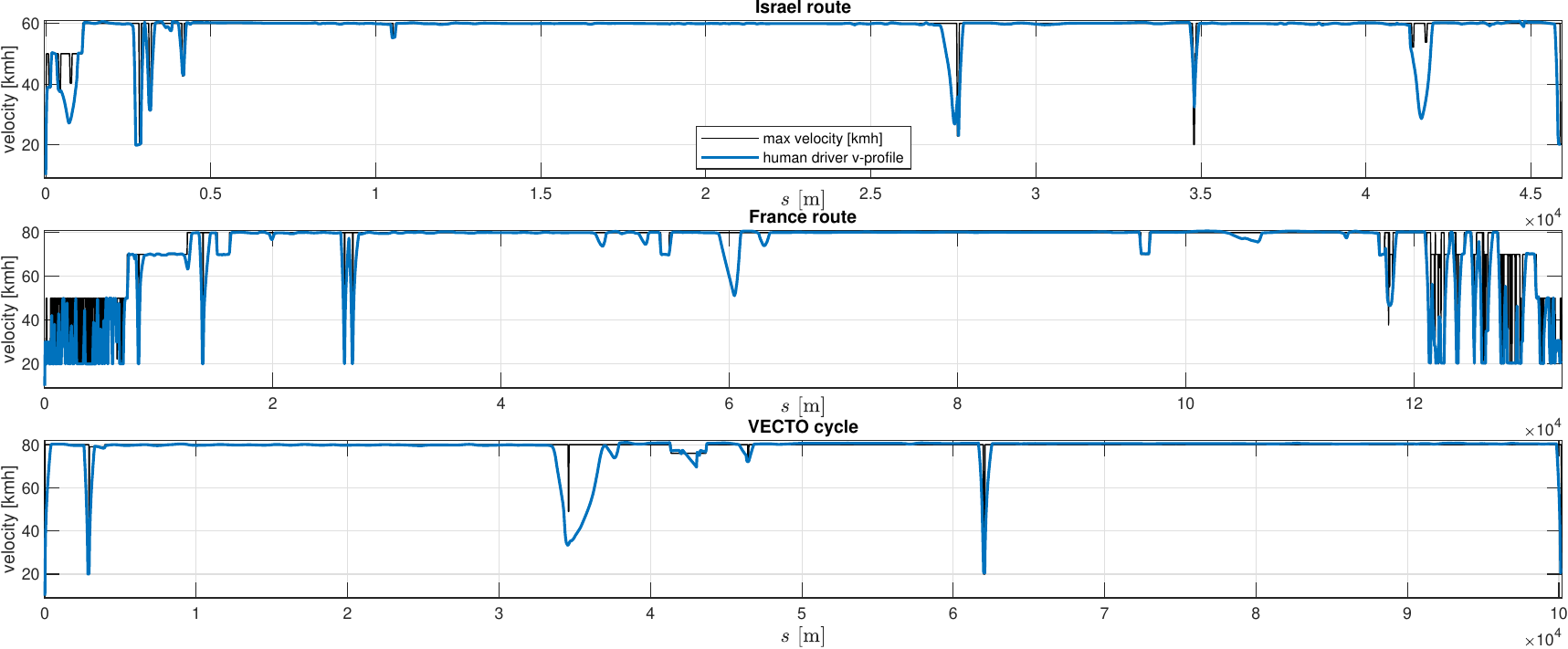}
    \caption{Simulated human driver velocity profiles. Respectively from top to bottom: Israel route, France route, VECTO long haul cycle.}
    \label{fig:humandriver}
\end{figure*}

To assess the relative fuel savings compared to existing methods, a comparison is made with the driving behavior of a (simulated) human driver. This human driving behavior forms a baseline reference and consists of a speed controller to track the maximum velocity as velocity reference and a previewing time in which the HDT can decelerate earlier in case the velocity reference decreases in the previewed horizon. For the longitudinal speed controller we use a PI controller with parameters $K_p=10000$ and $K_i=1$. The preview time is the equivalent of the anticipating behavior of a real human driver and, based on \cite{xing2017study}, is set to 
$$t_{prev} = 2.8+0.25(v_0-\text{min}(v_{m,prev})),$$ with $v_{m,prev}$ the maximum velocity profile of the previewed horizon. Next, we model the gear shift heuristic as in \cite{gearh} with a gear upshift if $\omega_e>2000$ and a downshift if $\omega_e<1000$. In addition, a low gear is kept in downhill situations to increase the braking torque \cite{5274169}. The baseline reference velocity profile for the Israel route is illustrated in Fig. \ref{fig:humandriver}. For this scenario, the total fuel consumption was 2621.5~[g] with a trip duration of 2886~[s].

\subsection{Simulation results: influence of the tuning parameters}\label{subsec:results}

In this section we analyze the influence of the tuning parameters $\phi$, $\beta$ and $N$ on the fuel consumption and time duration of the optimized profiles. For that, the routes are simulated for combinations of $\phi\in\{4,6,8,10,20,30,40,50,60,\\70,80,100\}$, $\beta\in\{1,5,10,20,50,100,$ $200\}$ and $N\in\{100,\\200,300,400\}$. 

\begin{figure}[t]
    \centering
    \includegraphics[width=0.9\linewidth]{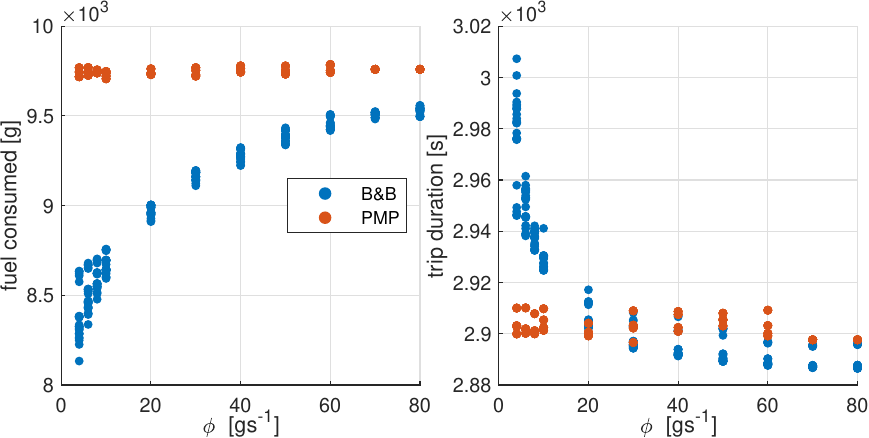}
    \caption{Total fuel consumption and trip duration for various simulations of the Israel route for the ED MPC Bnb and PMP solver.}
    \label{fig:phiIsrael}
\end{figure}

The influence of the parameter $\phi$ on fuel consumption and trip duration illustrated in Fig. \ref{fig:phiIsrael}. One can observe that for the optimal solution an increase in $\phi$ correlates with an increased fuel consumption and shorter trip duration; this behavior was consistent for the three routes. 
A higher $\phi$ increases the velocity and thereby the fuel consumption. This correlation trend is also stronger for the BnB solver. Additionally, a larger spread in data points is visible for low $\phi$ values, suggesting that the influence of $N$ and $\beta$ increases as $\phi$ decreases. However, 
no trend in the influence of $\beta$ was observed.

\begin{figure}[t]
    \centering
    \includegraphics[width=0.7\linewidth]{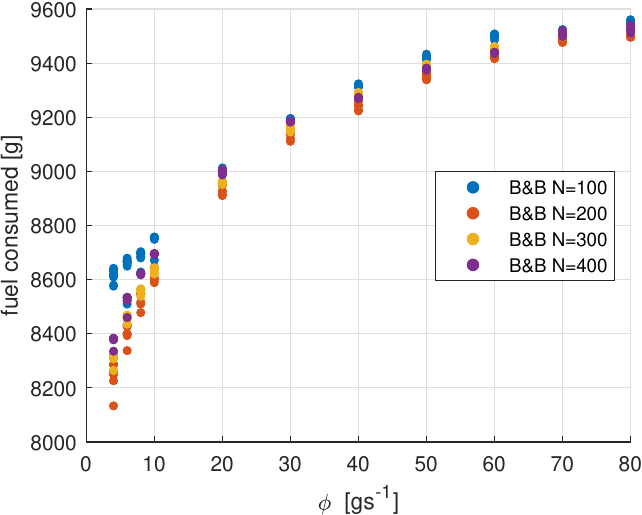}
    \caption{Influence of prediction horizon $N$ on the Israel route for different $\phi$. Observe that a larger $N$ does not always yield a lower fuel consumption.}
    \label{fig:Nstudy}
\end{figure}

Next we examined the influence of the prediction horizon $N$, as it significantly affects computation times (see Section \ref{ssec:comp_time}). From Fig. \ref{fig:Nstudy} we notice that for a larger $\phi$, the prediction horizon has a relatively small influence on fuel consumption. Therefore, using larger $\phi$-values enables the BnB MPC solver to also run on less powerful hardware.

On the other hand, a lower $\phi$ increases the influence of $N$ on fuel consumption, although a larger horizon does not necessarily  result in lower fuel savings and $N=200$ seems to be optimal. The main cause for this is that the average $v_{max}$--constraint (which is approximately followed) and the relative potential energy of the HDT are prediction horizon--dependent, such that the fuel consumption and trip duration do not increase proportionally with the horizon increase.

\subsection{Real--time constraint}\label{ssec:comp_time}

The algorithms were simulated in Matlab, which has no explicit code compilation step. Once compiled in C++, for instance, a 20x speedup can be achieved \cite{nazar}. In \cite{thomassen} the real--time constraint is quantified with a policy update of 1 [Hz] for a driver advisory system, resulting in a time constraint of 20 [s] on the optimization time in Matlab. The average optimization computation times of the routes are presented in Table \ref{tab:cpu}, along with the \textit{estimated} computation time in C++. One can observe that the constraint is met and that the computation time increases disproportionately with a horizon increase.

\begin{table}[h]
\centering
\caption{Average computation time of the optimization intervals in Matlab and C++. The C++ times are estimated and based on a 20x speedup \cite{nazar}. Values must remain under 20[s] and 1[s], respectively for Matlab and C++.}\label{tab:cpu}
\begin{tabular}{c|cccccc}
\hline
          & \multicolumn{6}{c}{Computation time {[}s{]}}                                                            \\ \cline{2-7} 
          & \multicolumn{2}{c|}{Israel route} & \multicolumn{2}{c|}{France route} & \multicolumn{2}{c}{VECTO cycle} \\ \cline{2-7} 
horizon $N$[-] & Matlab & \multicolumn{1}{c|}{C++} & Matlab & \multicolumn{1}{c|}{C++} & Matlab           & C++          \\ \hline
$N$=100       & 0.1729      & \multicolumn{1}{c|}{0.0086}   & 0.2590      & \multicolumn{1}{c|}{0.0130} & 0.2399 & 0.0120 \\
$N$=200       & 0.5202      & \multicolumn{1}{c|}{0.0260}   & 0.7364      & \multicolumn{1}{c|}{0.0368}  & 0.6598 &  0.0330 \\
$N$=300       & 1.0387     & \multicolumn{1}{c|}{0.0519}  &  1.3930     & \multicolumn{1}{c|}{0.0697}  & 1.2316  & 0.0616 \\
$N$=400 & 1.7609 & \multicolumn{1}{c|}{0.0880}  & 2.2655    & \multicolumn{1}{c|}{0.1133}  & 2.0182 & 0.1009  \\ \hline
\end{tabular}
\end{table}

\begin{figure*}[h]
    \centering
    \includegraphics[width=\linewidth]{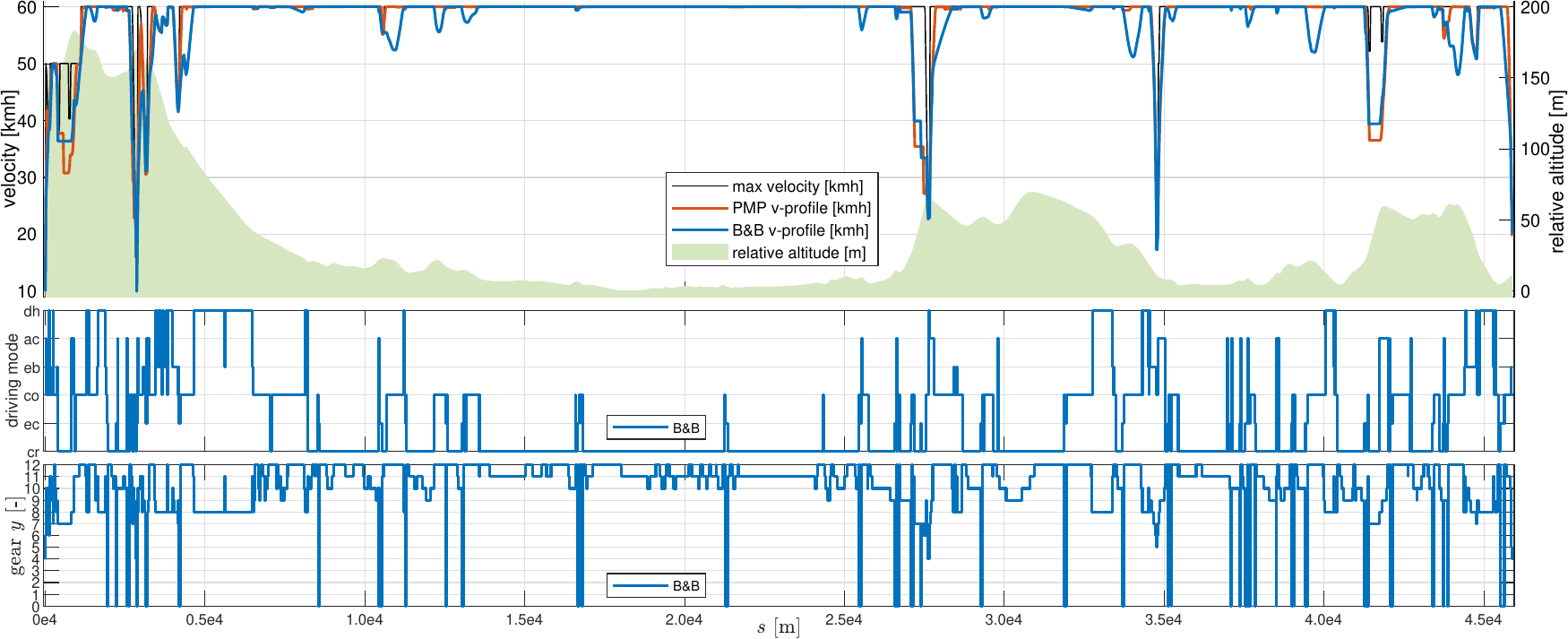}
    \caption{Israel route. Respectively from top to bottom: velocity--trajectories; driving mode; gear position. Simulation settings: $N=200$, $\phi=10$, $\beta=10$. Results of this particular simulation (total fuel consumed[g], trip duration[s]): PMP ED MPC: 2901[s],9722[g]; BnB ED MPC: 2927[s],8601[g]; Performance: BnB vs. PMP: 0.90\% time increase, 11.53\% fuel saved; BnB vs. baseline reference: 1.43 \% time increase, 31.85\% fuel saved.}
    \label{fig:fullroute1}
\end{figure*}

\subsection{Simulation results: final route simulation}
Based on the observations made in the previous sections and given that in practical applications, especially in delivery situations, only a certain in increase in trip duration is admissible, we now set our choice of parameters as follows. A maximum trip duration increase of 5\% with respect to the baseline reference trip duration is allowed, leading to a choice of $\phi=10$. Based on the trend in Fig. \ref{fig:Nstudy} and a trade--off with computational time, we also selected $N=200$. Finally, since for this range of parameters the final parameter $\beta$ has no clear influence, we used $\beta=10$.

The total fuel consumption, trip duration, and relative fuel savings for the three routes are presented in Table \ref{tab:relsave}. One can observe that the BnB solver consistently outperformed the PMP solution and the human driver baseline reference. In addition, it can also be observed in Table \ref{tab:relsave} that the highest fuel savings occurred on the Israel route, while this is also the shortest route (Table \ref{tab:routes}). It is therefore likely that extending the Israel route will yield similar fuel savings as the other two. The resulting velocity profile with corresponding driving mode and gear position for the Israel route is illustrated in Fig. \ref{fig:fullroute1}.

One can observe that the BnB and PMP velocity profiles are similar and relatively close to $v_{max}$. The main differences are mostly observed at road gradient changes and $v_{max}$ transients. Consider for instance the velocities around $11000$~[m]: around this point, the BnB velocity is lower than the PMP velocity and the altitude profile indicates a local maximum. This means that part of potential energy is temporarily converted into kinetic energy and that BnB is able to capture this advantage more efficiently.

This performance difference can be explained by two main differences: while PMP is only guaranteed to be locally optimal \cite{locallyOptimal} (by the the initial co-state guess) and focuses in matching the boundary constraints, the BnB algorithm is approximately globally optimal (from a driving modes perspective) and therefore finds a lower cost. 

Finally, one can observe that the driving modes and gear switching occur can rather frequently since it is not penalized in the cost function. For a more realistic outcome, a constraint on the number of switches per distance could be added.

\setlength{\tabcolsep}{0.2em}
\begin{table}[H]\caption{Total fuel consumption and trip duration of the baseline reference, and relative fuel savings of ED MPC BnB and PMP.}\label{tab:relsave}
\centering
\begin{tabular}{@{}c|cc|ccc@{}}
\hline
                      & \multicolumn{2}{c|}{baseline reference}                        & \multicolumn{3}{c}{fuel savings {[}\%{]}} \\ \cline{2-6} 
\multicolumn{1}{l|}{} & \multicolumn{1}{l}{Fuel con.} & \multicolumn{1}{l|}{Trip dur.} & BnB vs.      & PMP vs.    & BnB vs.     \\
route {[}-{]}         & [g]                       & [s]                        &b. ref.     & b. ref.          & PMP         \\ \hline
Israel route  & 12621.5                 & 2886.0                  & 31.85  &22.97 & 11.53                    \\
France route  & 42904.6                 & 7545.6                  &  23.18 & 9.95  & 14.87                  \\
VECTO cycle   & 27898.7                  & 4674.2                 & 18.56&12.03& 7.44                   \\ \hline
\end{tabular} 
\end{table}

\section{CONCLUSIONS}\label{sec:conclusion}
In this work, we proposed a dedicated MPC BnB solver for a driving modes ED
problem including slope and V2X preview information to meet real--time implementability constraints. We introduced two new driving modes for the inclusion of look--ahead slope information in the OCP. To make the solver real--time implementable, context--based node elimination was applied, a context--based heuristic in the lower bound was used and the solver was warm--started with a PMP solution. Then, we evaluated the results of the BnB solver, as well as a human driver model and a MPC PMP--based solution in a numerical simulation of three routes. In comparison with a human driver and a PMP solution, respectively, up to 25.8\% and 12.9\% fuel savings are achieved on average. The BnB solver also consistently outperforms the human driver and the MPC PMP-–based solution. These fuel savings were consistently achieved at the lowest time/fuel weight ratios in the cost function, but not necessarily at the largest prediction horizon. 

In future work, it can be investigated whether the developed driving modes method approached the optimal solution that is based on continuous control variables. Moreover, three things are expected to improve performance: 1) the heuristic can be improved using a feedforward artificial neural network; 2) the nodes can be branched in parallel; 3) the warm--started initial solution can be improved by also using the optimal driving mode sequence of the previous optimization interval. 



\bibliographystyle{IEEEtran}
\bibliography{references}

\newpage
\section*{APPENDIX}
In this section we present the results of the France route and VECTO cycle. The results of the Israel route are presented in Section \ref{sec:results}. 
\begin{figure*}[h!]
    \centering
    \includegraphics[width=\linewidth]{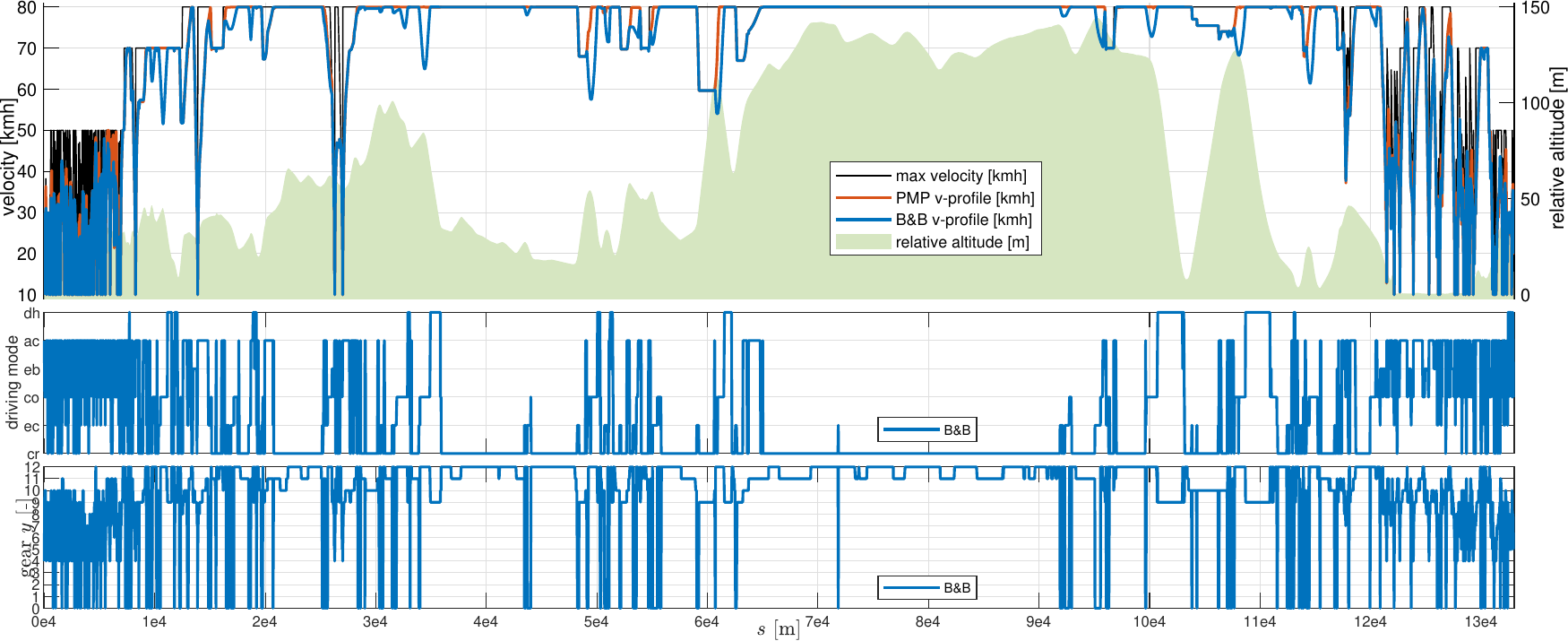}
    \caption{France route. Respectively from top to bottom: velocity-trajectories; driving mode; gear position. Simulation settings: $N=200$, $\phi=10$, $\beta=10$. Results of this particular simulation (total fuel consumed[g], trip duration[s]): PMP ED MPC: 7621[s],38909[g]; BnB ED MPC: 7706[s],35976[g]; Performance: BnB vs. PMP: 1.12\% time increase, 7.54\% fuel saved; BnB vs. baseline reference: 2.13\% time increase, 16.15\% fuel saved.}
    \label{fig:fullroute2}
\end{figure*}
\begin{figure*}[b!]
    \centering
    \includegraphics[width=\linewidth]{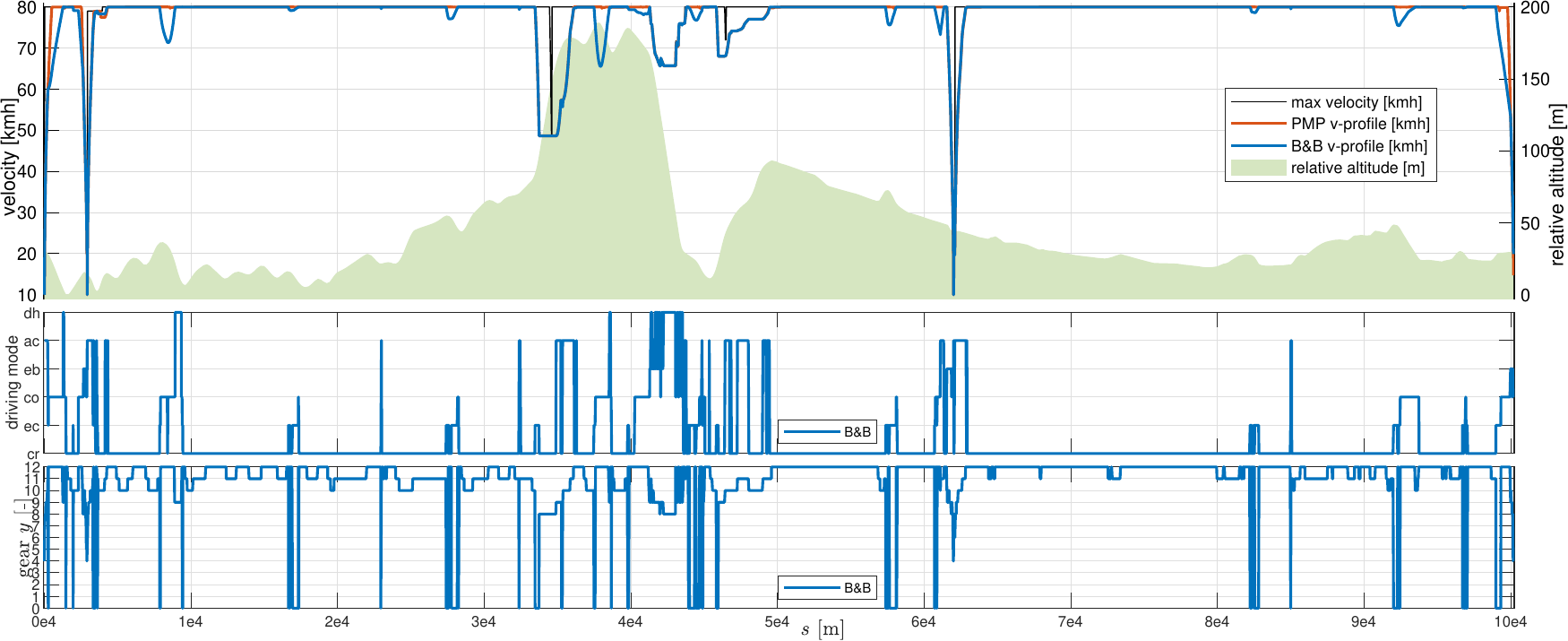}
    \caption{VECTO long haul cycle. Respectively from top to bottom: velocity-trajectories; driving mode; gear position. Simulation settings: $N=200$, $\phi=10$, $\beta=10$. Results of this particular simulation (total fuel consumed[g], trip duration[s]): PMP ED MPC:4686[s],24547[g]; BnB ED MPC: 4719[s],23671[g]; Performance: BnB vs. PMP: 0.70\% time increase, 3.57\% fuel saved; BnB vs. baseline reference: 0.96\% time increase, 15.15\% fuel saved.}
    \label{fig:fullroute3}
\end{figure*}

\begin{figure}[H]
    \centering
    \includegraphics[width=\linewidth]{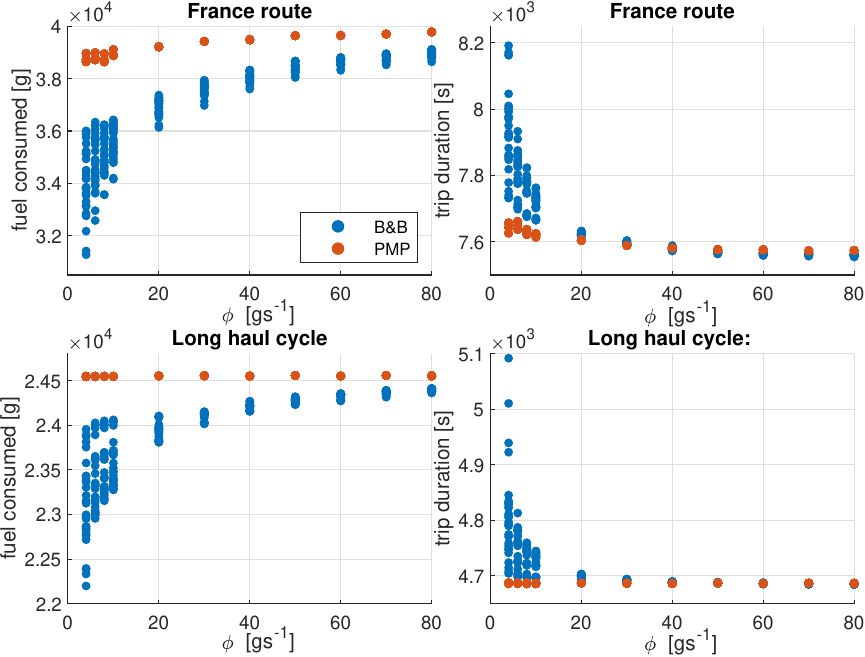}
    \caption{Total  fuel  consumption  and  trip  duration  for various simulations of the  France route (top) and VECTO cycle (bottom). This figure complements Fig. \ref{fig:phiIsrael}.  }
    \label{fig:phiall}
\end{figure}

\begin{figure}[H]
    \centering
    \includegraphics[width=\linewidth]{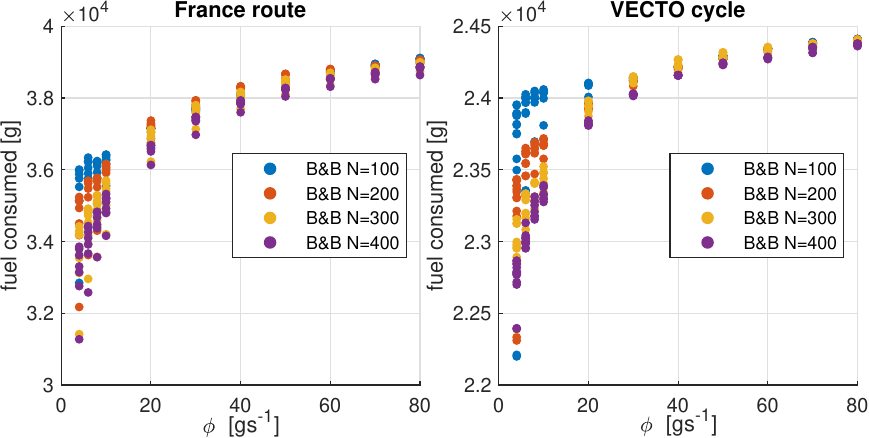}
    \caption{Influence of horizon $N$ on the France route (left) and VECTO cycle (right). This figure complements Fig. \ref{fig:Nstudy}.}
    \label{fig:NstudyAll}
\end{figure}


\end{document}